\documentclass[journal]{IEEEtran}
\usepackage{multicol}
\usepackage{lipsum,graphicx,subcaption}
\usepackage{float}
\usepackage{epstopdf}
\usepackage{ifthen}
\usepackage[latin1]{inputenc}

\usepackage{amssymb}
\usepackage[cmex10]{amsmath}
\usepackage{dcolumn}
\usepackage{fixltx2e}
\usepackage[nolist]{acronym}
\DeclareGraphicsExtensions{.pdf,.jpeg,.png,.eps}
\usepackage{booktabs}

\usepackage{comment}
\usepackage{color}

\usepackage{setspace}
\usepackage{mathrsfs}          
\usepackage{amsmath, amssymb, amsthm}
\usepackage{amsfonts}
\usepackage{mathrsfs}
\usepackage{multirow}        
\usepackage{framed}
\usepackage{psfrag}
\usepackage{graphicx}
\usepackage{units}            
\usepackage{algpseudocode} 

\newcolumntype{d}[1]{D{.}{\cdot}{#1} }

\newcommand{\mat}[1]{\mathbf{#1}}

\usepackage{bm}        
\newcommand{\LeftP} {\left\lbrace } 
\newcommand{\RightP}{\right\rbrace } 

\newcommand{\LeftPD} {\left( } 
\newcommand{\RightPD}{\right)} 

\newcommand{\compl}{\mathbb{C}}         
  
\newcommand{\ma}  [1]{ \bm{#1} }

\newcommand{\frob}[1]  {\left \| #1 \right\| _F}

\newcommand{\NORM}[1]  { \| #1 \|  }

\newcommand{\IndexM}[3]{\left[ #1\right]_{\LeftPD #2,#3 \RightPD } } 

\newcommand{\Vect}  [1] {\mathrm{vec}  \LeftP #1 \RightP }
\newcommand{\diag} [1] {\mathrm{diag} \LeftP #1 \RightP }

\newcommand{\unvec} [3] {\mathrm{unvec}_{#2 \times #3} \LeftP #1 \RightP }

\newcommand{\Circ} [2] {#1^{( #2)}}
\newcommand{\DFT} [1] {\ma{F}_{#1}} 
\newcommand{\Amat}  {\ma{A}} 
\newcommand{\mav}  [1]{ \bm{#1} }
\newcommand{\R}  [2]{ \ma{R}_{#1,#2} }
\newcommand{\U}  [2]{ \ma{U}_{#1,#2} }
\newcommand{\PI}  [2]{ \ma{\Pi}_{#1,#2} }
\newcommand{\V}  [3]{ \ma{V}^{(#3)}_{#1,#2} }
\newcommand{\Z}  [3]{ \ma{Z}^{(#3)}_{#1,#2} }

\begin{document}
\begin{acronym}
	\acro{OOB}{Out-of-Band }
	\acro{RRC}{Root-Raised Cosine}
	\acro{ISI}{inter-symbol-interference}
	\acro{ZF}{zero-forcing}
	\acro{MF}{matched filter}
	\acro{SINR}{signal-to-interference-plus-noise-ratio}
	\acro{SNR}{signal to noise ratio}
	\acro{FIR}{finite impulse repose }
	\acro{DFT}{discrete Fourier transform}
	\acro{RC}{raised-cosine}
	\acro{GFDM}{generalized frequency division multiplexing}
	\acro{ICI}{inter-carrier-interference}
	\acro{NEF}{noise-enhancement factor}
	\acro{FDE}{Frequency Domain Equalization}
	\acro{SVD}{singular-value decomposition}
	\acro{AWGN}{additive white Gaussian noise}
	\acro{DTFT}{discrete-time Fourier transform}
	\acro{OFDM}{orthogonal frequency division multiplexing}
	\acro{FFT}{fast Fourier transform}
	\acro{SIR}{signal-to-interference ratio}
	\acro{DZT}{discrete Zak transform}
\end{acronym}

\title{Optimal Radix-2 FFT Compatible Filters for GFDM}

\author{
\IEEEauthorblockN{Ahmad Nimr, Maximilian Matth\'{e}, Dan Zhang,  Gerhard Fettweis}\\
\IEEEauthorblockA{ Vodafone Chair Mobile Communication Systems, Technische Universit\"{a}t Dresden, Germany}\\
\IEEEauthorblockA{\small\texttt{\{first name.last name\}}}\IEEEauthorblockA{\small\texttt{@ifn.et.tu-dresden.de}}
}

\maketitle
\IEEEpeerreviewmaketitle
\begin{abstract}
	For a linear waveform, a finite condition number of the corresponding modulation matrix is necessary for the waveform to convey the message without ambiguity. Based on the Zak transform, this letter presents an analytical approach to compute the condition number of the modulation matrix for the multi-carrier waveform \ac{GFDM}. On top, we further propose a filter design that yields non-singular modulation matrices for an even number of subcarriers and subsymbols, which is not achievable for any previous work.  Such  new design has significant impact on implementation complexity, as the radix-2 FFT operations for conventional multicarrier waveforms can readily be employed for GFDM. Additionally, we analytically derive the optimal filter that minimizes the condition number. We further numerically evaluate the \ac{SIR} and \ac{NEF} for \ac{MF} and \ac{ZF} \ac{GFDM} receivers for such design, respectively.
\end{abstract}

\section{Introduction}
Among several waveform alternatives to \ac{OFDM} \cite{Banelli2014}, considerable research on detection algorithms, performance and low-complexity implementations has been conducted for \ac{GFDM}  \cite{GFDM}. \ac{GFDM}, as a non-orthogonal filtered multicarrier system with $K$ subcarriers employs circular filtering of $M$ subsymbols within each block to keep the signal confined within the block duration of $KM$ samples. Naturally, the choice of the pulse shaping filter strongly influences the system performance, as it controls system orthogonality and interference structure. In \cite{Gabor}, it is proved by means of the \ac{DZT} \cite{Bolcskei1997} of the transmit filter that the transmit signal becomes ambiguous and the \ac{GFDM} modulation matrix $\ma{A}$ is singular when a real-valued symmetric filter with even $M$ and $K$ is employed. The authors of \cite{singular} extended this result by showing that $\ma{A}$ has exactly one zero eigenvalue, suggesting that one GFDM block can mostly convey $(KM-1)$ data symbols. Hence, odd $M$ is commonly adopted  for data transmission in the literature. Even though several works on complexity reduction for \ac{GFDM} modulation and demodulation have been published \cite{FDE,Zhang2015}, odd $M$ forbids the $N$-point FFT to be implemented solely by energy-efficient radix-2 based processing. This also narrows the design space of GFDM as a flexible waveform generator \cite{Flexible_GFDM}.

In this paper we propose a filter design for \ac{GFDM} to supports even values for both $M$ and $K$,  particularly when they are power-of-two
. To this end, we introduce a fractional shift in the sampling of the continuous frequency response of conventional basis filters, such as \ac{RC} filter,  to allow both even-valued and odd-valued $M,K$ to be derived from the same filter response. 
As a function of this shift, a closed-form expression for the condition number of $\Amat$ is provided and the optimal shift for both even and odd $M$, $K$ in terms of the minimal condition number is derived. To verify the design we evaluate the \ac{SIR} for the \ac{MF} receiver and the \ac{NEF} of the \ac{ZF} receiver.

section{GFDM Modulation Matrix Decomposition}\label{sec:GFDM review}
One \ac{GFDM} block conveys the data symbols $\{d_{k,m}\}$ via $K$ subcarriers and $M$ subsymbols, yielding $N=KM$ samples. The $n$th one as the entry $n$ of $\ma{x}\in\compl^{N\times1}$ equals~\cite{GFDM}
\begin{equation}
	[\ma{x}]_{n} = \sum\limits_{k=0}^{K-1}\sum\limits_{m=0}^{M-1}d_{k,m}g[\left<n-mK\right>_N]e^{j2\pi\frac{k}{K}n}, \label{eq: GFDM sum form}
\end{equation}
where $g[n]$ denotes the pulse shaping filter and corresponds to the entry $n$ of $\ma{g}\in\compl^{N\times 1}$. With the $N\times N$ modulation matrix $\Amat$ constructed as $\IndexM{\Amat}{n}{k+mK} = g[\left<n-mK\right>_N]e^{j2\pi\frac{k}{K}n}$, Eq. (\ref{eq: GFDM sum form}) can be formed as $\mav{x}=\ma{A}\ma{d}$, where $[\ma{d}]_{k+mK}=d_{k,m}$.
\subsection{Decomposition of $\Amat$}
For a $L\times{}Q$ matrix $\ma{X}$, let $\mav{x}=\text{vec}_{L,Q}(\ma{X})$ and $\text{unvec}_{L,Q}(\mav{x})$ denote the vectorization operation and its inverse. Let $\DFT{N}$ be the $N$-point \ac{DFT} matrix with elements $\IndexM{\DFT{N}}{i}{j} = e^{-j2\pi\frac{ij}{N}}$. Let the unitary matrix $\U{L}{Q}$ be
\begin{equation}
	\U{L}{Q} = \frac{1}{\sqrt{Q}}\ma{I}_L\otimes\DFT{Q},\label{eq: U matrix}
\end{equation}
where $\otimes$ is the Kronecker product. Let $\PI{L}{Q}\in \Re^{LQ\times LQ}$ be the permutation matrix  that fulfills for any $L\times Q$ matrix~$\ma{X}$
\begin{equation}
	\text{vec}(\ma{X}^T)=\PI{L}{Q}\text{vec}(\ma{X}).
\end{equation}
The $(Q,L)$ \ac{DZT} \cite{Bolcskei1997} \mbox{$\ma{Z}\mav{x}=\left(\DFT{Q}\otimes\ma{I}_{L}\right)\mav{x}$}, \mbox{$\mav{x}\in \compl^{QL\times 1}$} can be written as a matrix $ \Z{Q}{L}{\mav{x}}\in \compl^{Q\times L}$ with
\begin{align}
	\Z{Q}{L}{\mav{x}}&= \DFT{Q}\V{Q}{L}{\mav{x}} = \tilde{\ma{V}}_{Q,L}^{(\mav{x})}, \label{eq: ZAK}\\
	\text{where }\quad\V{Q}{L}{\mav{x}}&= \left(\unvec{\mav{x}}{L}{Q} \right)^T. \label{eq: reshape x}
\end{align}
Resorting to the \ac{DZT} of $\ma{g}$ and $\tilde{\ma{g}}=\DFT{N}\ma{g}$, we can factorize $\Amat$ into two forms
\begin{align}
	\Amat &= \underbrace{\PI{K}{M}^T\U{K}{M}^H}_{\ma{U}^{(g)}}\ma{\Lambda}^{(g)}\underbrace{\U{K}{M}\PI{K}{M} \U{M}{K}^H}_{\ma{V}^{(g)H}}, \label{eq: A-TD}\\
	&= \underbrace{\frac{\DFT{N}^H}{\sqrt{N}}\PI{M}{K}^T\U{M}{K}^H}_{\ma{U}^{(\tilde{g})}}\ma{\Lambda}^{(\tilde{g})}\underbrace{\U{M}{K}\PI{M}{K} \U{K}{M}\PI{M}{K}}_{\ma{V}^{(\tilde{g})H}}, \label{eq: A-FD}
\end{align}
where $\ma{U}^{(\tilde{g})}$ and $\ma{V}^{(\tilde{g})}$ are unitary matrices and
\begin{align}
	\ma{\Lambda}^{(g)} &= \diag{\text{vec}_{M,K}\LeftP\sqrt{K}\Z{M}{K}{\mav{g}}}\RightP, \label{eq: D-TD}\\
	\ma{\Lambda}^{(\tilde{g})} &= \diag{\text{vec}_{M,K}\LeftP\frac{1}{\sqrt{K}}\Z{K}{M}{\tilde{\mav{g}}}}\RightP, \label{eq: D-FD}
\end{align}
where $\ma{\Lambda}^{(g)}$ and $\ma{\Lambda}^{(\tilde{g})}$ contain the
\ac{DZT} of $\mav{g}$ and $\mav{\tilde{g}}$, respectively. As a result, properties of $\mat{A}$ 
are dictated by $\ma{\Lambda}^{(\tilde{g})}$ or $\ma{\Lambda}^{({g})}$. 
In this paper we focus on pulse shapes that are sparse in frequency, hence we employ $\ma{\Lambda}^{(\tilde{g})}$ for the subsequent analysis.

\subsection{Performance indicators}
Define the short-hand notation $z_{k,m} = \IndexM{\Z{K}{M}{\tilde{g}}}{k}{m}$. Then $\sigma^2_{k,m} =|z_{k,m}|^2$ correspond to the squared singular values of $\Amat$ scaled by $K$. The conditional number of $\Amat$ is given by \cite{SVD}
\begin{equation}
	\mathrel{cond}(\Amat) = \frac{\max_{k,m}\{\sigma_{k,m} \}}{\min_{k,m}\{\sigma_{k,m}\}} = \frac{\sigma_{max}}{\sigma_{min}}. \label{eq: cond}
\end{equation}
Considering the received signal in AWGN channel \cite{Receiver} the \ac{NEF} of \ac{ZF}
and the \ac{SIR} of the \ac{MF} receiver can be written as
\begin{equation}
	\begin{split}
		\mbox{NEF} &=\frac{1}{N^2}\frob{\Amat}^2\frob{\Amat^{-1}}^2 =  \frac{1}{N^2} \frob{\ma{\Lambda}^{(\tilde{g})}}^2\frob{\ma{\Lambda}^{(\tilde{g})-1}}^2\\
		&= \frac{1}{N^2}\left(\sum\limits_{k,m}\sigma^2_{k,m}\right)\left(\sum\limits_{k,m}\frac{1}{\sigma^2_{k,m}}\right),
	\end{split}\label{eq: NE}
\end{equation}
\begin{equation}
	\begin{split}
		\mbox{SIR} &= \frac{1}{N}\frob{ \frac{\Amat^H\Amat}{\NORM{\ma{g}}^2 }- \ma{I}_N}^2= \frac{1}{N}\frob{ \frac{\ma{\Lambda}^{(\tilde{g})}\ma{\Lambda}^{(\tilde{g})H}}{\NORM{\ma{g}}^2}-\ma{I}_N}^2\\
		&= \frac{1}{N}\sum\limits_{k,m}\left(\frac{\sigma_{k,m}^2}{{\frac{1}{N}\sum\limits_{k,m}\sigma_{k,m}^2}}-1\right)^2.
	\end{split} \label{eq: MF factor}
\end{equation}

\section{GFDM pulse shaping filter design}\label{sec:pulse shape design}
The conventional pulse shaping filter design for the \ac{GFDM} is to let $\mav{g}=\DFT{N}^H\tilde{\mav{g}}$ with $[\tilde{\mav{g}}]_n=H(\tfrac{n}{N})$, where $H(\nu)$ stands for the \ac{DTFT} of a pre-selected basis filter $h[n]$ that is of practical interests, e.g., \ac{RC} or \ac{RRC}. Here, $\nu$ is the normalized frequency and thus the period of $H(\nu)$ is equal to $1$. With such design of $\mav{g}$, it has been shown in \cite{Gabor} that $\Amat$ becomes singular for even $M,K$ and a real symmetric filter $h[n]$. This is caused by \mbox{$\IndexM{\ma{Z}_{K,M}^{(\mav{\tilde{g}})}}{k}{m}=(\mathcal{Z}H(\nu))(\tfrac{k}{K}, \tfrac{m}{M})$}, where $(\mathcal{Z}H(\nu))(f,t)$ denotes the discrete-time Zak transform of $H(\nu)$ and for any real symmetric  filter we know $(\mathcal{Z}H(\nu))(\tfrac{1}{2},\tfrac{1}{2})=0$. The requirement of odd $M$ or $K$ impedes an efficient implementation in terms of low-complexity \mbox{radix-$2$} FFT operations. In the sequel, we propose a novel design approach that overcomes this restriction for any basis filter $h[n]$ fulfilling the following conditions
\begin{enumerate}
	\item $h[n]$ is real-valued, i.e. $H(\nu) = H^*(1-\nu)=H^*(-\nu).$
	\item $H(\nu)$ spans two subcarriers within each period, i.e. $H(\nu) = 0$, $\forall\nu \in [\frac{1}{K}, \frac{1}{2}]$.
	\item $|H(\nu)|$ is decreasing from $1$ to $0$ for $\nu \in [0, \frac{1}{K}]$.
\end{enumerate}
We start from noting
\begin{align}
	|(\mathcal{Z}H(\nu-\eta))(f,t)|=|(\mathcal{Z}H(\nu))(f-\eta,t)|,
\end{align}
namely, shifting the frequency response of a filter also shifts the frequency coordinate of its Zak transform \cite{Bolcskei1997}. Hence, shifting $H(\nu)$ can help us avoid to sample the zero in $(\mathcal{Z}H(\nu))$ for even $M,K$.
Accordingly, the samples of $\tilde{\mav{g}}$ are defined as
\begin{equation}
	\hspace{-0.1cm}[\tilde{\ma{g}}]_n(\lambda) \hspace{-0.1cm}=\hspace{-0.1cm} \left\lbrace\hspace{-0.25cm}
	\begin{array}{cc}
		H\left(\frac{n+\lambda}{N}\right),& \hspace{-0.2cm}0\leq n < M-\lambda \\
		H^*\left(\frac{N-n-\lambda}{N}\right),&\hspace{-0.2cm} N -M-\lambda < n \leq N-1\\
		0,&\hspace{-0.2cm} \mbox{otherwise}
	\end{array}
	\hspace{-0.25cm}\right\rbrace,
\end{equation}
for $\lambda\in[0,1[$. $\tilde{\ma{g}}$ can be reshaped as in \eqref{eq: reshape x} to
\begin{equation}
	\hspace{-0.1cm}\IndexM{\V{K}{M}{\tilde{\ma{g}}}(\lambda)}{k}{m} = \left\lbrace \hspace{-0.15cm}\begin{array}{cc}
		H\left(\frac{m+\lambda}{N}\right),& k=0\\
		H^*\left(\frac{M-m-\lambda}{N}\right),& k=K-1\\
		0,& \mbox{elsewhere}
	\end{array}
	\hspace{-0.15cm}\right\rbrace
	.\label{reshape  subcarriers}
\end{equation}
Applying \ac{DFT} according to \eqref{eq: ZAK}, we get
\begin{equation}
	z_{k,m}(\lambda) = H\left(\frac{m+\lambda}{N}\right)+H^*\left(\frac{M-m-\lambda}{N}\right) e^{j2\pi \frac{k}{K}}\label{eq: Zak gf}.
\end{equation}
Due to the symmetry of $H(\nu)$, we have
\begin{equation}
	\begin{split}
		z_{k,m}(1-\lambda) &= z_{k,M-1-m}^*(\lambda)e^{j2\pi\frac{k}{K}},\\ \sigma^2_{k,m}(1-\lambda) &= \sigma^2_{k,M-1-m}(\lambda).
	\end{split} \label{eq: sing}
\end{equation}
Hence, all results regarding conditional number, \ac{NEF} and \ac{SIR} are symmetric around $\lambda = 0.5$. Moreover, Eq. \eqref{eq: Zak gf} shows that $z_{k,m}(1+\lambda) = z_{k,m+1}(\lambda)$. Hence, it suffices to study the case $0\leq \lambda \leq 0.5$. Additionally, we focus on $K = 2^x$ for $x>1$.

To obtain closed-forms of the condition number of $\Amat$, we subsequently focus on two particular families of $H(\nu)$, namely well-localized filters that fulfill the \ac{ISI}-free criterion without or with matched filtering.
\subsection{ISI free without matched filter}\label{sub: 3.2.A}
In this case $H(\nu)$ additionally satisfies
\begin{equation}
	\sum\limits_{k=0}^{K-1}H\left(\nu -\frac{k}{K}\right) = 1. \label{eq: ISI free criteria}
\end{equation}
From the symmetry and limited band  of $H(\nu)$ it follows
\begin{equation}
	H(\nu) + H^*\left(\frac{1}{K}-\nu\right) = 1,~ \forall \nu \in [0, \frac{1}{K}] \label{eq: property form ISI free }
\end{equation}
and $H\left(\frac{m+\lambda}{N}\right)+H^*\left(\frac{M-m-\lambda}{N}\right) = 1$.
Also, there exists a function $f(\nu) = r(\nu)e^{j\phi(\nu)}$ with $f(\nu) = -f^*\left(\frac{1}{K}-\nu\right)$ and
\begin{equation}
	H(\nu) = \frac{1}{2}\left(1+f(\nu)\right), \forall \nu \in [0, \frac{1}{K}]. \label{eq: ISI free}
\end{equation}
Let us assume a real-valued $f(\nu)$, i.e. $\phi(\nu) = 0$ and $f(\nu) = r(\nu)$\footnote{Complex $f(\nu)$ as in Xia-filters \cite{xiafamily} is treated in the following section.}.  Due to the constraint of decreasing amplitude $H(\nu)$, $r(\nu)$ must be decreasing from $1$ to $-1$ for $\nu \in [0,\frac{1}{K}]$. Based on \eqref{eq: property form ISI free } and \eqref{eq: ISI free}  we get
\begin{equation}
	\begin{split}
		\sigma^2_{A_{k,m}}(\lambda)
		&=\frac{(1+f^2_m(\lambda)}{2} +\frac{(1-f^2_m(\lambda))}{2}\cos\left(2\pi\frac{k}{K}\right)
	\end{split}, \label{eq: singular ISI free}
\end{equation}
where $f_m(\lambda) = f(\frac{m+\lambda}{N}) = 2H\left(\frac{m+\lambda}{N}\right)-1$.
The singular values are symmetric with respect to $k$, and decreasing with $k = 0,\cdots, \frac{K}{2}$.
Therefore, $\sigma^2_{A_{0, m}}(\lambda) =1 $ and $\sigma^2_{A_{\frac{K}{2}, m}}(\lambda) = f_m^2(\lambda)$ are the maximum and minimum singular value with respect to $k$, respectively. Therefore, $\sigma^2_{A_{max}}(\lambda) = 1$, because $f_m^2(\lambda)\leq 1 $, and $\sigma^2_{A_{min}}(\lambda)$ is obtained from $\min_m\{f_m(\lambda)^2\}$.
Since $f(\nu)$ is decreasing and antisymmetric around $\frac{1}{2K}$, $f(\nu)^2$ is  decreasing $\forall \nu \in [0,~\frac{1}{2K}]$ and increasing $\forall \nu \in [\frac{1}{2K},~ \frac{1}{K}]$. As a result,
when $M$ is even, $0\leq \lambda\leq 0.5$, $	\sigma_{A_{min}}^2$ is obtained at $m = M/2$, and when $M$ is odd, it is obtained at $m = (M-1)/2$. Therefore,
\begin{equation}
	\sigma_{A_{min}}^2(\lambda) =
	f^2\left(\frac{1}{2K}+\frac{S(\lambda)}{2N}\right). \label{eq: sig_min ISI free}
\end{equation}
\begin{equation}
	\text{where}\quad S(\lambda) = \left\lbrace
	\begin{array}{cc}
		2\lambda,& M \mbox{ is even} \\
		1-2\lambda,& M \mbox{ is odd}
	\end{array}
	\right\rbrace.
\end{equation}
From the increasing/decreasing intervals of $f^2(\nu)$, $\sigma_{A_{min}}^2(\lambda)$ increases with $0\leq \lambda \leq 0.5$ for even $M$ and decreases when $M$ is odd.
Hence, the condition number can be expressed as
\begin{equation}
	\mathrm{cond}(\Amat_A)(\lambda) =
	\frac{1}{\left|f\left(\frac{1}{2K}+\frac{S(\lambda)}{2N}\right)\right|}. \label{eq: cond ISI free}
\end{equation}
Similarly, $\mathrm{cond}(\Amat_A)(\lambda)$ is decreasing for even $M$ and increasing for odd $M$. Hence, the best condition of $\Amat$ is attained at $\lambda = 0.5$ for even $M$ and $\lambda = 0$ for odd $M$.
\subsection{ISI free after matched filtering}
A filter $H(\nu)$ is ISI-free after matched filtering if
\begin{equation}
	\sum\limits_{k=0}^{K-1}\left|H\left(\nu -\frac{k}{K}\right)\right|^2 = 1.
\end{equation}
By exploiting the symmetry and band limit, we get
\begin{equation}
	|H(\nu)|^2 + \left|H^*\left(\frac{1}{K}-\nu\right)\right|^2 = 1,~ \forall \nu \in [0, \frac{1}{K}], \label{eq: half free ISI Property}
\end{equation}
and hence $\left|H\left(\frac{m+\lambda}{N}\right)\right|^2+\left|H^*\left(\frac{M-m-\lambda}{N}\right)\right|^2 = 1$.
Furthermore, there exists a real-valued function $f(\nu) = -f\left(\frac{1}{K}-\nu\right)$, which is decreasing from $1$ to $-1$ in the interval $\nu \in [0, \frac{1}{K}]$  with
\begin{equation}
	|H(\nu)|^2 = \frac{1}{2}(1+f(\nu)), \forall \nu \in [0, \frac{1}{K}]. \label{eq: half property form ISI free }
\end{equation}
Adding an (arbitrary) phase $\phi(\nu)$ yields the original $H(\nu)$ by
\begin{equation}
	H(\nu) = e^{j\phi(\nu)}\sqrt{\frac{1}{2}(1+f(\nu))}, \forall \nu \in [0, \frac{1}{K}]. \label{eq: ISI free matched}
\end{equation}
Using \eqref{eq: half free ISI Property} and \eqref{eq: ISI free matched},
\begin{equation}
	\begin{split}
		H\left(\frac{m+\lambda}{N}\right) &=   e^{j\phi_{a,m}(\lambda)}\sqrt{\frac{1}{2}(1+f_m(\lambda))}, \\
		H^*\left(\frac{M-m-\lambda}{N}\right) &=   e^{j\phi_{b,m}(\lambda)}\sqrt{\frac{1}{2}(1-f_m(\lambda))}.
	\end{split}
\end{equation}
where
$\phi^a_m(\lambda) = \phi\left(\frac{m+\lambda}{N}\right)$,
$\phi^b_m(\lambda) = -\phi\left(\frac{M-m-\lambda}{N}\right)$, and
$f_m(\lambda)= f\left(\frac{m+\lambda}{N}\right)$. As special cases, we study the phase in the form $\phi(\nu) = - \phi\left(\frac{1}{K}-\nu\right) + \beta \frac{\pi}{2}$,$~ \beta = 0,~1,~2,~3$.  Then $e^{j\phi_{a,m}(\lambda)} = j^{\beta}e^{j\phi_{b,m}(\lambda)}$. No \ac{ISI} with and without \ac{MF}, as the Xia filters \cite{xiafamily} provide, is obtained with $f(\nu) = \cos(2\phi(\nu))$ and $\beta = 2$ or, equally, $\phi(\nu)=\tfrac{1}{2}\mathrm{acos}(f(\nu)$. 
From \eqref{eq: Zak gf}, we get
\begin{equation}
	\begin{split}
		\sigma^2_{B_{k,m}}(\lambda) &=  1+ \sqrt{1-f_m^2(\lambda)}\cos\left(2\pi\frac{k - \beta \frac{K}{4}}{K}\right ). \label{eq: singular matched}
	\end{split}
\end{equation}

The maximum singular value with respect to $k$ is located at $k_{max} = \beta \frac{K}{4}$ and the minimum one at $k_{min} = (\beta+2\mod{4}) \frac{K}{4}$. This requires that $K$ is a multiple of $4$ for $\beta = 1, 3$.
\begin{equation}
	\begin{split}
		\sigma^2_{B_{k_{max}, m}}(\lambda)& = 1+\sqrt{1-f_m^2(\lambda)},\\
		\sigma^2_{B_{k_{min}, m}}(\lambda)& = 1-\sqrt{1-f_m^2(\lambda)}.
	\end{split}
\end{equation}
\begin{figure*}[h]
	\centering
	\captionbox{Conditional number.\label{fig1}}[.32\linewidth][c]{%
		\includegraphics[width=.35\linewidth]{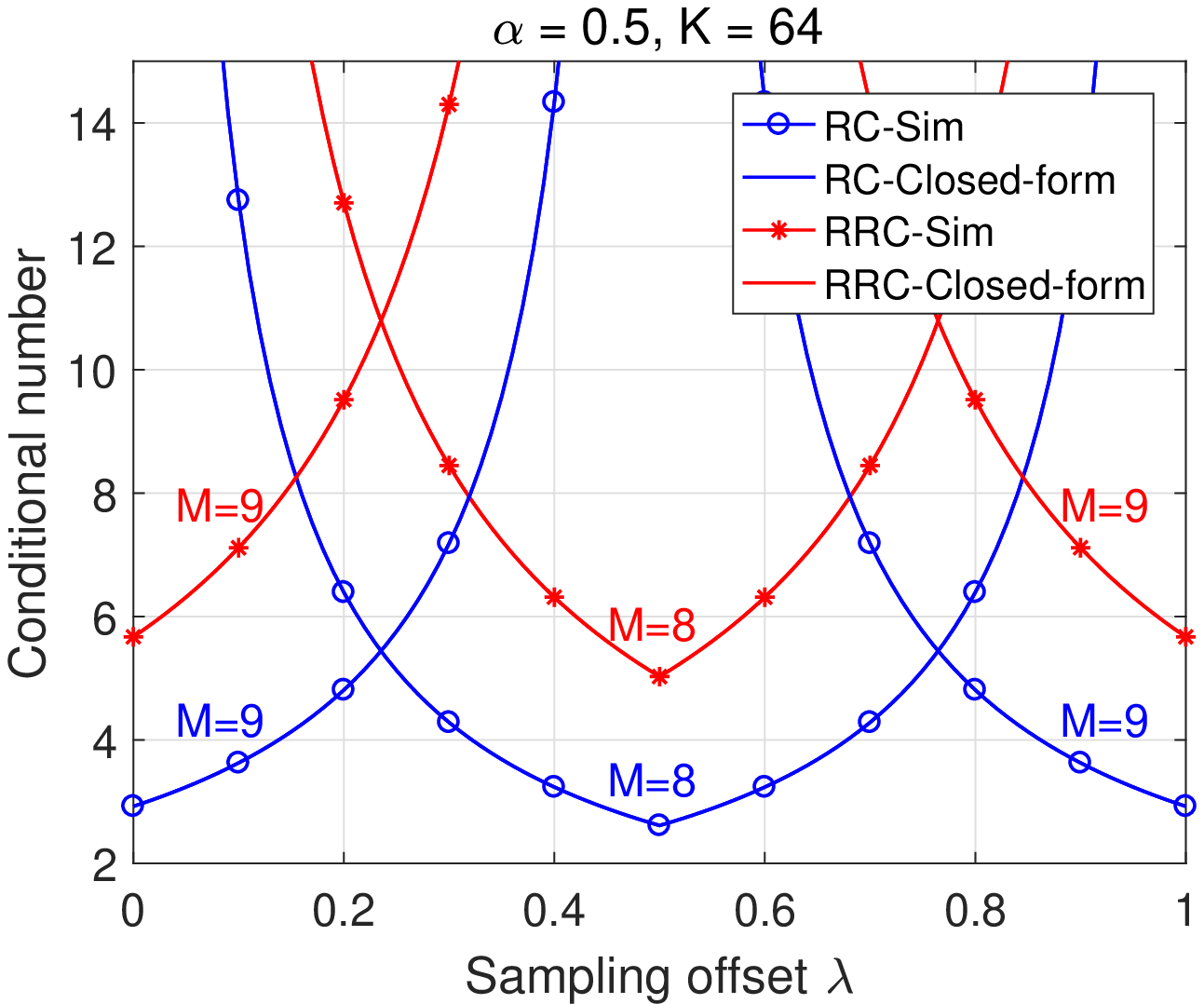}}
	\captionbox{Noise enhancment Factor. \label{fig2}}[.32\linewidth][c]{%
		\includegraphics[width=0.35\linewidth]{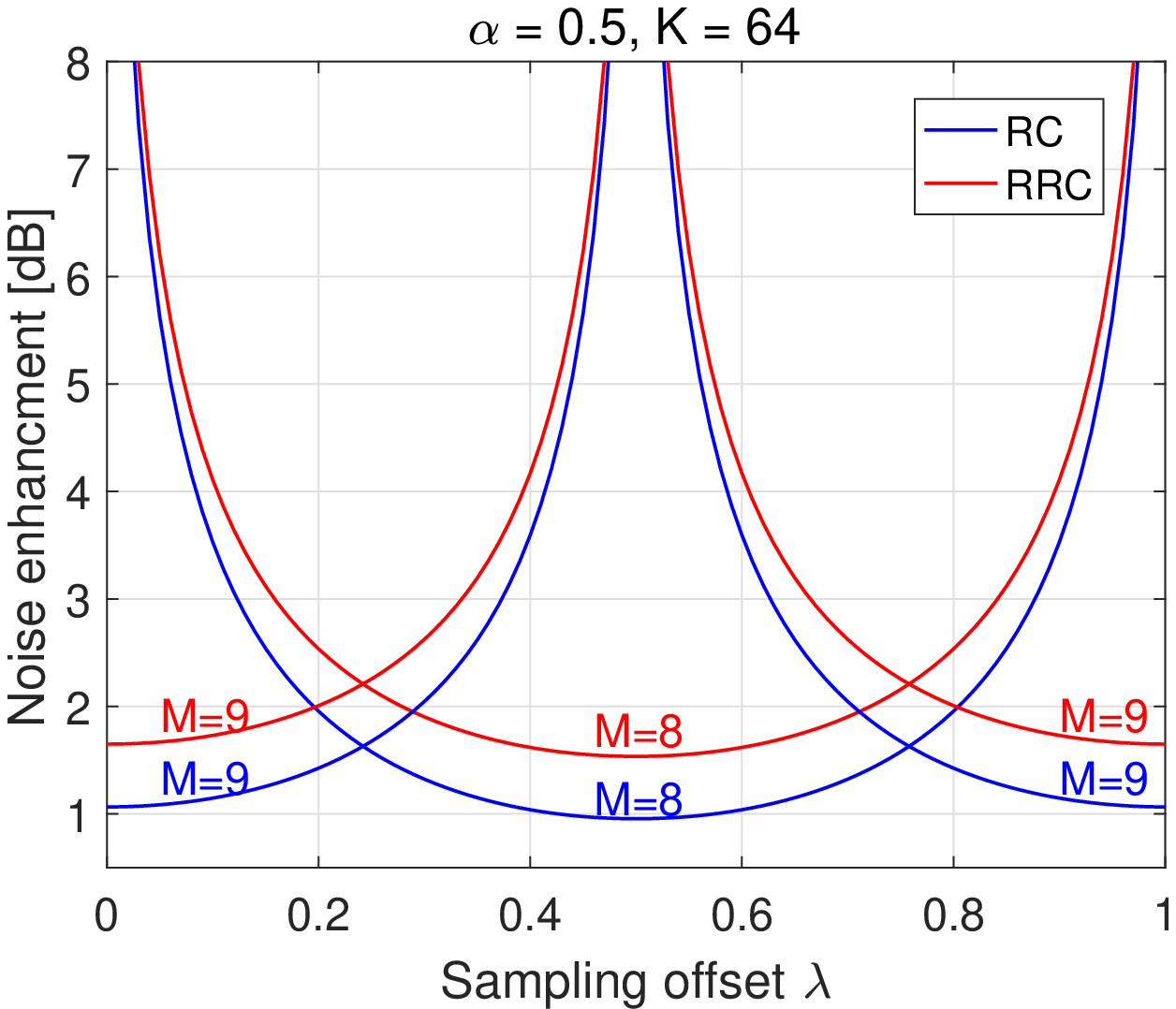}}
	\captionbox{\ac{NEF} and \ac{SIR} for optimal
		$\lambda$.\label{fig3}}[.32\linewidth][c]{%
		\includegraphics[width=0.35\linewidth]{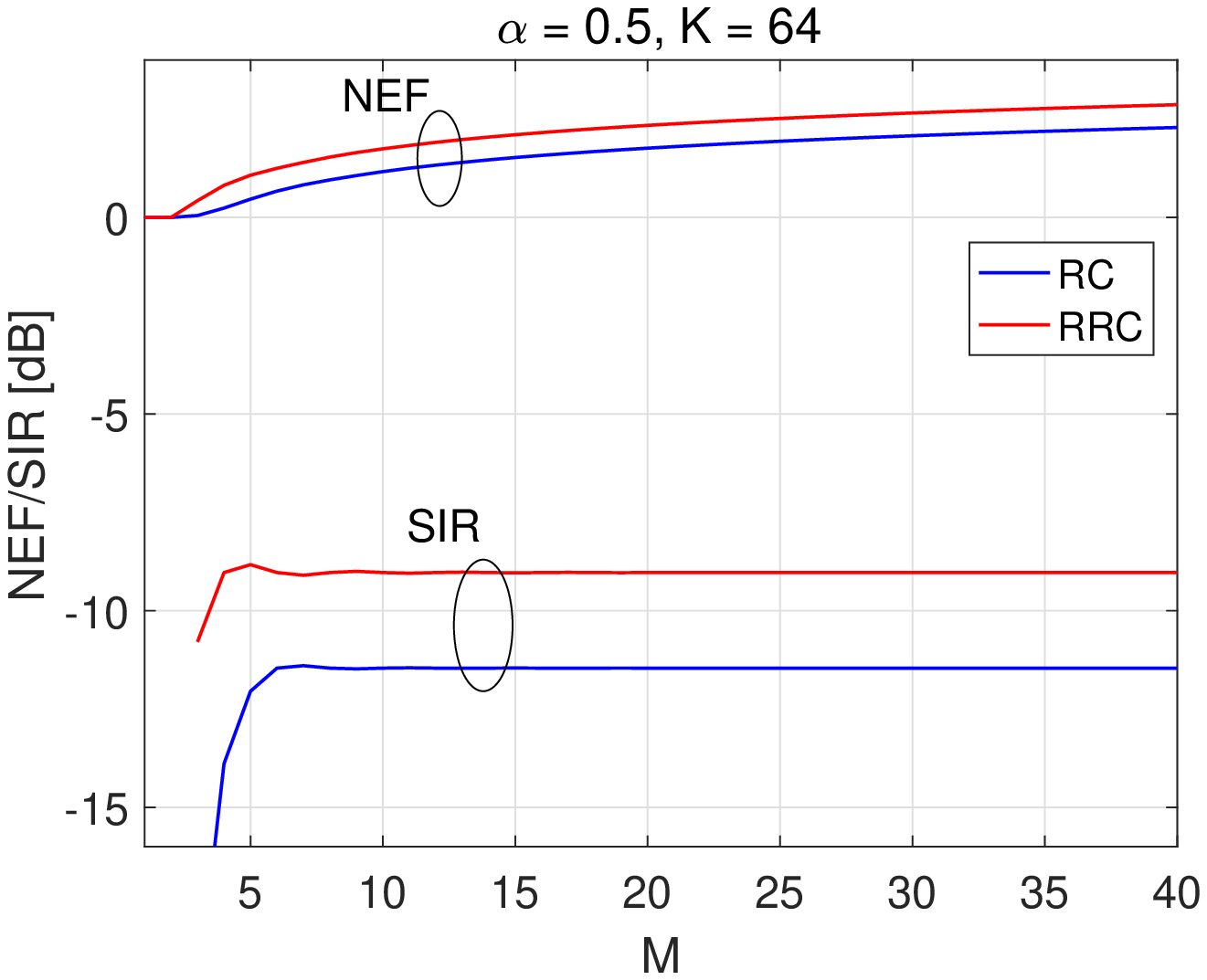}}
\end{figure*}
Following the same argument as previously, based on the properties of  $f(\nu)$, both $\sigma^2_{B_{min}}(\lambda)$ and $\sigma^2_{B_{max}}(\lambda)$ are obtained at $m = M/2$ for even $M$ and $m=\frac{M-1}{2}$ and for odd $M$. Thus,
\begin{equation}
	\begin{split}
		\sigma_{B_{max}}^2(\lambda) &=
		1+\sqrt{1-f^2\left(\frac{1}{2K}+\frac{S(\lambda)}{2N}\right)},\\
		\sigma_{B_{min}}^2(\lambda) &=
		1-\sqrt{1-f^2\left(\frac{1}{2K}+\frac{S(\lambda)}{2N}\right)},
	\end{split}
	\label{eq: sig_min ISI matched}
\end{equation}
and the conditional number can then be written as
\begin{equation}
	\mathrm{cond}(\Amat_B)(\lambda) =
	\frac{\left|f\left(\frac{1}{2K}+\frac{S(\lambda)}{2N}\right)\right|}{1-\sqrt{1-f^2\left(\frac{1}{2K}+\frac{S(\lambda)}{2N}\right)}}.
	\label{eq: cond Matched}
\end{equation}
$\mathrm{cond}(\Amat_B)(\lambda)$ is decreasing for even $M$ and increasing for odd $M$ with $\lambda \in [0,0.5]$.
When using the same function $f(\nu)$ in cases A and B, we notice that   $\sigma^2_{B_{max}}(\lambda)\geq 1 = \sigma^2_{A_{max}}$ and $\sigma^2_{B_{min}}\leq f_{m_{max}}^2(\lambda) = \sigma^2_{A_{min}}$, and hence
\begin{equation}
	\mathrm{cond}(\Amat_A)(\lambda) \leq \mathrm{cond}(\Amat_B)(\lambda), \label{eq: compare cond}
\end{equation}
proving that the condition number is smaller when using an ISI-free filter, compared to using its square root, which has been numerically shown in \cite{Matthe14}.

\section{Numerical example}\label{sec:numerical example }
In this section, we study the family of prototype filters with
roll-off factor $\alpha$, being obtained with the generator  function
\begin{equation}
	f(\nu) = \left\lbrace
	\begin{array}{cc}
		1,& 0\leq \nu \leq \frac{1-\alpha}{2K}\\
		f^a\left(\frac{2K}{\alpha}[\nu-\frac{1}{2K}]\right), &\frac{1-\alpha}{2K} < \nu \leq \frac{1+\alpha}{2K}\\
		-1,& \frac{1+\alpha}{2K} < \nu \leq \frac{1}{K} \label{eq: f_alpha}
	\end{array}
	\right\rbrace.
\end{equation}
$f^a$ is real-valued anti-symmetric ($f^a(x) = f^a(-x)$), and decreasing from $1$ to $-1$ for  $x\in[-1,1]$. Therefore,  $f(\nu) = -f(\frac{1}{K}-\nu)$. Hence,  $f(\nu)$ can construct pulse shapes according to \eqref{eq: ISI free} or \eqref{eq: ISI free matched}. From \eqref{eq: cond Matched}, \eqref{eq: cond ISI free} and using \eqref{eq: f_alpha}, we find that
for $M\alpha \leq  S(\lambda)$, $\mathrm{cond}(\Amat_A) =\mathrm{cond}(\Amat_B) =1 $. For $ S(\lambda) \leq M\alpha$,
\begin{equation}
	\begin{split}
		\mathrm{cond}(\Amat_A)(\lambda) &=
		\frac{1}{\left|f^a\left(\frac{S(\lambda)}{\alpha M}\right)\right|},\\
		\mathrm{cond}(\Amat_B)(\lambda) &=
		\frac{\left|f^a\left(\frac{S(\lambda)}{\alpha M}\right)\right|}{1-\sqrt{1-f^{a2}(\frac{S(\lambda)}{\alpha M})}}.
	\end{split}
	\label{eq: cond study case}
\end{equation}
The condition number is independent of $K$ and, based on the properties of $f^a$, increases with $\alpha M$. 
As a particular example, \ac{RC} and \ac{RRC} use the function $f^a (x) = -\sin(\frac{\pi}{2}x)$. Replacing in \eqref{eq: cond study case} we get,
\begin{equation}
	\begin{split}
		\mathrm{cond}(\Amat_{\text{RC}})(\lambda) &=
		\left(\sin\left(\frac{\pi}{2}\frac{S(\lambda)}{\alpha M}\right)\right)^{-1},\\
		\mathrm{cond}(\Amat_{\text{RRC}})(\lambda) &=
		\left(\tan\left(\frac{\pi}{4}\frac{S(\lambda)}{\alpha M}\right)\right)^{-1}.
	\end{split}\label{eq: RC RCC cond}
\end{equation}
Fig. \ref{fig1} shows the condition number of $\Amat$ for different sampling shift $\lambda$, and validates the closed-form expressions \eqref{eq: RC RCC cond} numerically. As shown, $\lambda=0$ is optimal for odd $M$ and $\lambda=\tfrac{1}{2}$ for even $M$ when $K$ is also even. In addition, as proved in \eqref{eq: RC RCC cond}, using \ac{RC} yields a better conditioned $\Amat$ than \ac{RRC}. Furthermore, numerically obtained values for \ac{NEF} as shown in Fig. \ref{fig2} behave similarly as the condition number. This can be explained by the influence of the smaller singular value on the noise enhancement. In both cases, the condition number as well as the smallest singular value depend on $\left|f^a\left(\frac{S(\lambda)}{\alpha M}\right)\right| = \sin\left(\frac{\pi}{2}\frac{S(\lambda)}{\alpha M}\right)$.
Considering the optimum $\lambda$, Fig. \ref{fig3} illustrates \ac{NEF} and \ac{SIR} with different $M$. The proper choice of $\lambda$ with respect to $M$ preserves the trend of \ac{NEF} which increases with $M$. On the other hand, \ac{SIR} is independent of $M$ when $M$ is big enough. In fact, \ac{SIR} approaches the interference value that can be directly obtained from  $\text{SIR}=2\int_{\tfrac{1}{2K}}^{\tfrac{1}{2}}|H(\nu)|^2d\nu$, which is independent of $\lambda$ and $K$ but depends on $\alpha$.


\section{Conclusion}\label{sec:conclusions}
For the waveform \ac{GFDM}, the condition number of its modulation matrix is fully characterized by the adopted pulse shaping filter. In this letter, we observed that a frequency-domain shift of the frequency response of the pulse shaping filter can yield a change in the condition number. By deriving a closed-form expression of the condition number, we can find the optimal shift that minimizes the condition number for \ac{GFDM} modulation. This yields a filter design that permits GFDM to have an arbitrary numbers of subcarriers and subsymbols per subcarrier, in particular power-of-two values become possible. We numerically verified the obtained closed-form expression and computed the \ac{NEF} and \ac{SIR} with respect to \ac{ZF} and \ac{MF} receivers in an \ac{AWGN} channel, indicating that an optimal condition number yields also optimal \ac{NEF} values.

\section*{Acknowledgment}
The work presented in this paper has been performed in the framework of the SATURN project with contract no. 100235995 funded by the Europaischer Fonds f\"ur regionale Entwicklung (EFRE) and by the Federal Ministry of Education and Research within the programme "Twenty20 - Partnership for Innovation" under contract 03ZZ0505B - "fast wireless".

\bibliographystyle{IEEEtran}
\bibliography{IEEEabrv,references}
\appendix
\section{Appendix}\label{sec:append}
Let $\ma{S}\in \compl^{QL\times QL}$ be a block circulant matrix generated from diagonal matrices, such that
\begin{equation}
	\ma{S}=\left[
	\begin{array}{cccc}
		\ma{S}_0 &\ma{S}_{Q-1}&\cdots& \ma{S}_1\\
		\vdots & \vdots& &\vdots\\
		\ma{S}_{Q-1} &  \cdots &&\ma{S}_0
	\end{array}
	\right], \label{eq: block circulant}
\end{equation}
where
$\ma{S}_q = \diag{\mav{v}_q} \in \compl^{L\times L}$, and $\ma{v}_q $ is the $q$-th column of a matrix $\ma{V}\in \compl^{L\times Q}$, i.e. $\mav{v}_q = \IndexM{\ma{V}}{:}{q}$, then \cite{Circuilant}
\begin{align}
	\ma{S} &= \PI{L}{Q}^T\U{L}{Q}^H\ma{\Lambda}\U{L}{Q}\PI{L}{Q}. \label{eq:block diag decomposition}
\end{align} 
\begin{equation}
	\text{where,     }\ma{\Lambda} = \diag{\Vect{\DFT{Q}\ma{V}^T}}.
\end{equation} 
Using the notations $\Circ{\mav{x}}{i} = \mav{x}[<n-i>_N]$ and defining the repletion matrix $\R{L}{Q} \in \Re^{LQ\times L}$, $\R{L}{Q} = \ma{1}_L\otimes\ma{I}_Q$
, the transmitted \ac{GFDM} block in  \eqref{eq: GFDM sum form} can be expressed in the following vector form,
\begin{equation}
	\begin{split}
		\mav{x} &= \sum\limits_{k=0}^{K-1}\sum\limits_{m=0}^{M-1}d_{k,m}\diag{\Circ{\mav{g}}{mK}}\R{M}{K}\IndexM{\DFT{K}^H}{:}{k},\\
		&= \sum\limits_{m=0}^{M-1}\sqrt{K}\diag{\Circ{\mav{g}}{mK}}\R{M}{K}\frac{1}{\sqrt{K}}\DFT{K}^H\mav{d}_m\\
		&= \ma{S}^{(M)} \U{M}{K}^H \Vect{\ma{D}}= \Amat \cdot  \Vect{\ma{D}}.
	\end{split}
\end{equation}
Here, $\Amat = \ma{S}^{(M)} \U{M}{K}^H$ and
\begin{equation}
	\frac{\ma{S}^{(M)}}{\sqrt{K}} = \left[\diag{\Circ{\mav{g}}{0K}}\R{M}{K},~\cdots,~ \diag{\Circ{\mav{g}}{(M-1)K}}\R{M}{K}\right]
\end{equation}
is block circular matrix as in \eqref{eq: block circulant}, with  $\ma{S}^{(M)}_m = \diag{\sqrt{K}\IndexM{\V{K}{M}{\mav{g}}}{:}{m}}$. From  \eqref{eq:block diag decomposition}, we get
\begin{align}
	\ma{S}^{(M)} &= \PI{K}{M}^T\U{K}{M}^H\ma{\Lambda}^{(g)}\U{K}{M}\PI{K}{M},\\
	\ma{\Lambda}^{(g)} &= \sqrt{K}\diag{\Vect{\DFT{M}\left(\V{K}{M}{\mav{g}}\right)^T}}.\label{eq: SM}
\end{align}
As a result we get $\Amat$ defined in \eqref{eq: A-TD}. 

The $N$-FFT of \eqref{eq: GFDM sum form} results in 
\begin{equation}
	\tilde{\ma{x}}[n] = \sum\limits_{k=0}^{K-1}\sum\limits_{m=0}^{M-1}d_{k,m}\tilde{g}[<n-kM>_N]e^{-j2\pi\frac{m}{M}n}.
\end{equation}
Following similar steps we get
\begin{equation}
	\tilde{\mav{x}} = \DFT{N}\Amat\cdot \Vect{\ma{D}}.
\end{equation}
Here $\DFT{N}\Amat = \ma{S}^{(K)} \U{K}{M}\PI{M}{K}$ and
\begin{equation}
	\frac{\ma{S}^{(K)}}{\sqrt{M}} = \left[
	\begin{array}{c}
		\diag{\Circ{\tilde{\mav{g}}}{0M}}\R{K}{M},~\cdots,~ \diag{\Circ{\tilde{\mav{g}}}{(K-1)M}}\R{K}{M}
	\end{array}
	\right].
\end{equation}
By replacing $\ma{S}^{(K)}$ as in \eqref{eq: SM}, then
\begin{equation}
	\tfrac{1}{\sqrt{N}}\DFT{N}\Amat = \PI{M}{K}^T\U{M}{K}^H{\ma{\Lambda}^{(\tilde{g})}}\U{M}{K}\PI{M}{K} \U{K}{M}\PI{M}{K}.
\end{equation}
\begin{equation}
	\ma{\Lambda}^{(\tilde{g})} = \tfrac{1}{\sqrt{K}}\diag{\Vect{\DFT{K}\left(\V{M}{K}{\tilde{\mav{g}}}\right)^T}}.
\end{equation}
And finally by multiplying whit $\tfrac{1}{\sqrt{N}}\DFT{N}$, we get \eqref{eq: A-FD}.

\end{document}